\documentclass[aps,prd,twocolumn,groupedaddress,nofootinbib]{revtex4-1}

\usepackage{dsfont}
\usepackage{natbib}
\usepackage{amssymb}
\usepackage{graphicx}
\usepackage{amsmath}
\usepackage{url}
\usepackage{bm}
\usepackage{epstopdf}
\usepackage{color}
\usepackage{enumitem}
\usepackage[colorlinks=true,breaklinks]{hyperref}
\usepackage{mathrsfs}

\newcommand{\di}{\partial}
\newcommand{\f}{\mathscr{F}}

\def\be {\begin{equation}}

\def\ee  {\end{equation}}

\def\bea {\begin{eqnarray}}

\def\eea {\end{eqnarray}}

\begin{document}

\title{A big bounce, slow-roll inflation and dark energy from conformal gravity}

\author{Jack Gegenberg}
\email{geg@unb.ca}
\affiliation{Department of Mathematics and Statistics, University of New Brunswick, Fredericton, NB, Canada E3B 5A3}

\author{Shohreh Rahmati}
\email{shohreh.rahmati@unb.ca}
\affiliation{Department of Mathematics and Statistics, University of New Brunswick, Fredericton, NB, Canada E3B 5A3}

\author{Sanjeev S.\ Seahra}
\email{sseahra@unb.ca}
\affiliation{Department of Mathematics and Statistics, University of New Brunswick, Fredericton, NB, Canada E3B 5A3}

\date{\today}

\begin{abstract}

We examine the cosmological sector of a gauge theory of gravity based on the SO(4,2) conformal group of Minkowski space.  We allow for conventional matter coupled to the spacetime metric as well as matter coupled to the field that gauges special conformal transformations.  An effective vacuum energy appears as an integration constant, and this allows us to recover the late time acceleration of the universe.  Furthermore, gravitational fields sourced by ordinary cosmological matter (i.e.\ dust and radiation) are significantly weakened in the very early universe, which has the effect of replacing the big bang with a big bounce.  Finally, we find that this bounce is followed by a period of nearly-exponential slow roll inflation that can last long enough to explain the large scale homogeneity of the cosmic microwave background.

\end{abstract}

\maketitle

\section{Introduction}

In order to reconcile Einstein's general relativity with observational facts, modern cosmology incorporates a number of elements that are difficult to justify from fundamental physics:  For example, dark matter is required to describe the clustering of stars and galaxies, dark energy is required to explain the late time acceleration of the universe, and an inflationary mechanism in the early universe is needed to explain both the large scale homogeneity of the cosmic microwave background as well as the origin of primordial fluctuations.  

The simplest framework that incorporates all of these elements has become known as the concordance model of cosmology: namely, $\Lambda$CDM with single field inflation.  In this paradigm, one assumes the existence of cold dark matter (CDM) that is not part of the standard model of particle physics, a cosmological constant $\Lambda$ that is added by hand to the Einstein-Hilbert action of general relativity, and a scalar inflaton field with an appropriate potential to drive nearly de Sitter (dS) inflation for a finite period in the early universe.  (We note that none of these axioms of standard cosmology are \emph{inconsistent} with general relativity.)  Even after observational evidence is accounted for via these mechanisms, theoretical challenges remain for concordance cosmology.  For example, if inflation is finite the classical equations of motion for general relativity imply that the universe started with a big bang.  (It should be mentioned that while this singular initial state is conceptually unappealing, it is not actually in direct conflict with observations.)

In an attempt to avoid some of the more \emph{ad hoc} elements of concordance cosmology, many authors have considered the possibility that general relativity is not the correct theory of gravitation.  Some of the oldest modified gravity theories have attempted to provide a non-particle explanation of dark matter, but such models can have difficulty accounting for the clustering of galaxies and weak gravitational lensing \cite{lrr-2012-10}.  Alternative gravity theories have also been proposed to explain the late time acceleration of the universe \cite{Dvali:2004ph, Clifton:2011jh}, as well as early time inflationary acceleration \cite{Starobinsky:1980te}.  Quantum corrections to general relativity have been used to tame the big bang singularity; for example, in loop quantum cosmology semiclassical equations of motion yield a big bounce instead of a big bang \cite{Bojowald:2001xe}.

In this paper, we study the cosmological implications of a modified gravity model that simultaneously addresses the issues of the initial singularity, the mechanism driving inflation, and the late time acceleration of the universe.  Our model belongs to the class of gauge theories of gravity \cite{hayshir, ivnied, Utiyama:1956sy, kibble,Wheeler:1991ff,Hazboun:2013lra,Wheeler:2013ora,Gegenberg:2015gma} in which the central object is a gauge potential analogous to the gauge potentials of particle physics.  The action functional is taken to be quadratic in the field strength of the gauge potential, just as in conventional Yang-Mills theory.  Geometric quantities, such as the metric and connection, are defined as functions of the gauge potential.  This ultimately leads to a metric theory of gravity.  We take the gauge group to be the conformal group of Minkowski spacetime SO(4,2), and the resulting theory is invariant under local conformal (Weyl) transformations.  Non-cosmological aspects of the SO(4,2) gauge gravity model have been studied in \cite{Wheeler:1991ff,Hazboun:2013lra,Wheeler:2013ora,Gegenberg:2015gma}.

A ubiquitous feature of gauge-gravity theories is manifolds with non-vanishing torsion.  When models based on the Poincar\'e \cite{Shie:2008ms} or de Sitter groups \cite{Huang:2008ij, Huang:2009sk} were applied in cosmology, it was found that nonzero torsion can drive late time acceleration.  Actually, in the de Sitter case nonzero torsion is a necessary condition for the existence of non-radiation cosmological matter.  It should also be noted that there are other non-gauge gravity models where torsion is responsible for singularity avoidance and inflation in the early universe \cite{Poplawski:2010kb,Poplawski:2012ab,Poplawski:2011jz,Desai:2015haa}.  However, while the SO(4,2) theory considered here allows for nonzero torsion, in this work we find that it is not required to reconcile the model with observations, explain dark energy, or to alter early universe dynamics.  However, we do require a population of matter that couples directly to the fields gauging the generators of special conformal transformations.

In section \ref{sec:model} we present the action of our model, and write down the equations of motion assuming vanishing torsion and matter which couples to the metric as well as fields gauging special conformal transformations.  In section \ref{sec:friedmann}, we specialize to homogeneous and isotropic spacetimes and write down the Friedmann equation for the model.  In section \ref{sec:dynamics}, we discuss solutions of the Friedmann equation and demonstrate the existence of a bounce, slow-roll inflation and late time acceleration.  In section \ref{sec:discussion}, we summarize and discuss our results.

\section{Gravitational Yang-Mills theory}\label{sec:model}

We consider the gravity theory described in \cite{Gegenberg:2015gma}.  The reader can find a more complete description of the model in that paper; here, we focus on a subsector of the theory obtained by making a number of simplifying assumptions.  

Our model is based on the SO(4,2) conformal group of Minkowski spacetime, which is the largest group of transformations that leaves null geodesics invariant. We begin with the so(4,2)-Lie algebra-valued vector potential
\begin{equation}
\mathbf{A}_{\alpha} = A^{A}_{\alpha} \mathbf{J}_{A} =   e^{a}_{\alpha} \mathbf{P}_{a} + l^{a}_{\alpha} \mathbf{K}_{a} + \omega_{\alpha}^{ab} \mathbf{J}_{ab} + q_{\alpha}
\mathbf{D},
\end{equation}
where the $\mathbf{J}_A=\left\{\mathbf{P}_a,\mathbf{K}_a,\mathbf{J}_{ab},\mathbf{D}\right\}$ are the generators of the Lie algebra and $\alpha = 0 \ldots 3$ is a spacetime index.  The components of the associated field strength $\mathbf{F}_{\alpha\beta} = F^{A}_{\alpha\beta} \mathbf{J}_{A}$ are given by
\begin{equation} 	
F^{A}_{\alpha\beta} = \di_{\alpha} A^{A}_{\beta} - \di_{\beta} A^{A}_{\alpha} +
f^{A}{}_{BC} A^{B}_{\alpha} A^{C}_{\beta},
\end{equation}
with the structure constants defined by $[\mathbf{J}_{A},\mathbf{J}_{B} ] = f^{C}{}_{AB} \mathbf{J}_{C}$.

We identify various components of $\mathbf{A}_{\alpha}$ in the $\mathbf{J}_{A}$ basis with geometric quantities in a 4-dimensional manifold $M$ with Lorentzian metric $g_{\alpha\beta}$ and affine connection $\Gamma^{\alpha}{}_{\beta\delta}$.  In particular, we take $e^{a}_{\alpha}$ as the components of an  orthonormal frame fields on $M$, with $\omega^{ab}_{\alpha}$ as the associated connection one-forms.  Hence, the metric and connection are given by:
\begin{equation}
\label{eq:tetrad postulate} 	g_{\alpha\beta} = \eta_{ab} e^{a}_{\alpha} e^{b}_{\beta}, \quad \Gamma^{\gamma}{}_{\alpha\beta} =
e^{\gamma}_{a}(\di_{\alpha} e_{\beta}^{a}  + \omega^{ac}_{\alpha} e_{c\beta}).
\end{equation}
In these expressions, lowercase Greek and Latin indices are raised and lowered with $g_{\alpha\beta}$ and $\eta_{ab}$, respectively.  The curvature one-forms are anti-symmetric in their frame indices $\omega^{(ab)}_{\alpha} = 0$, from which it follows that the affine connection is metric-compatible \cite{Carroll:2004st}:
\begin{equation} 	
0 = \nabla_{\alpha} g_{\beta\gamma},
\end{equation}
where $\nabla_{\alpha}$ is the derivative operator defined by $\Gamma^{\alpha}{}_{\beta\delta}$.  The
Riemann curvature and torsion tensors of $M$ are given by:
\begin{align} \nonumber
R^{\mu\nu}{}_{\alpha\beta} & = e^{\mu}_{a} e^{\nu}_{b} (d\omega^{ab} + \omega^{ac} \wedge \omega_{c}{}^{b})_{\alpha\beta}, \\ T^{\alpha}{}_{\beta\gamma} &= e^{\alpha}_{a} (de^{a} + \omega^{ac}\wedge e_{c})_{\beta\gamma}.\label{eq:torsion def}
\end{align}
Note that in this model, it is not necessary to assume $T^{\alpha}{}_{\beta\gamma} = 2\Gamma^{\alpha}{}_{[\beta\gamma]}=0$; however, we will concentrate on the vanishing torsion case in this paper.

The action functional of the model is
\be 	
S = - \frac{1}{2g^{2}_\text{YM}} \int d^{4}x \sqrt{-g} g^{\alpha\mu} g^{\beta\nu} h_{AB} F^{A}_{\alpha\beta} F^{B}_{\mu\nu} + S_\text{m},\label{eq:action}
\ee
where $h_{AB} = f^{M}{}_{AN} f^{N}{}_{BM}$ is the Cartan-Killing metric on so(4,2). The non-trivial components of $h_{AB}$ are:
\begin{gather}
h_{a\bar{b}} = h_{\bar{a}b}=-2\eta_{ab}, \quad h_{14,14} = 2, \nonumber \\ h_{[ab][cd]} = h_{[cd][ab]}=-4\eta_{a[c}\eta_{d]b}.\label{eq:killing metric}
\end{gather}
The notation here is that $a,\bar{a}=0,1,2,3$ denote components in the direction of translations $\mathbf{P}_a$ and special conformal transformations $\mathbf{K}_a$, respectively.  The six indices $[ab]$ consist of $[12],[23],[31],[01],[02],[03]$ and denote directions along  the distinct non-zero generators $\mathbf{J}_{ab}$ of Lorentz transformations.  Finally, the index $14$ denotes the component in the direction of the generator $\mathbf{D}$ of dilatations. We view (\ref{eq:action}) and (\ref{eq:killing metric}) as the defining relationships for our model.

The action (\ref{eq:action}) is manifestly diffeomorphism invariant, and as demonstrated in \cite{Gegenberg:2015gma}, it is invariant under local gauge transformations described by an eleven parameter subgroup of SO(4,2) with generators $\left\{\mathbf{K}_a,\mathbf{J}_{ab},\mathbf{D}\right\}$.  The behaviour of the gauge potential under these infinitesimal gauge transformations is:
\begin{gather}
	A^{A}_{\alpha} \mapsto A^{A}_{\alpha} +  \partial_{\alpha}\epsilon^{A}+ f^{A}{}_{BC} A^{B}_{\alpha} \epsilon^{C}, \nonumber \\ \epsilon^{A} \mathbf{J}_{A} =\lambda^{a} \mathbf{K}_{a} + \Lambda^{ab} \mathbf{J}_{ab} + \Omega \mathbf{D}.
\end{gather}
In particular, the component of the gauge potential in the direction of $\mathbf{D}$ transforms as:
\begin{equation}
	\delta q_{\alpha} = \di_{\alpha}\Omega + \tfrac{1}{2} \lambda_{\alpha}.\label{eq:gaugepres}
\end{equation}
It is obvious that we can impose the gauge condition $q_{\alpha} = 0$ via a simple series of gauge transformations of the form $\epsilon^{A}\mathbf{J}_{A}=\lambda^{a}\mathbf{K}_{a}$. This gauge condition is preserved under the gauge transformation generated by:
\begin{equation}\label{eq:gauge 2}
	\epsilon^{A} \mathbf{J}_{A} = -2 e^{a\alpha} \di_{\alpha} \Omega \, \mathbf{K}_{a} + \Lambda^{ab} \mathbf{J}_{ab} + \Omega \mathbf{D}.
\end{equation}
Under this class of restricted gauge transformations the metric transforms as:
\begin{equation}
\delta g_{\alpha\beta} = \Omega g_{\alpha\beta}.
\end{equation}
That is, the model is invariant under local conformal (Weyl) transformations.  For the rest of this paper we will work in the $q_{\alpha} = 0$ gauge.

While $q_{\alpha} = 0$ is a gauge choice and may be imposed without loss of generality, we will also enforce a number of additional conditions that are actually physically restrictive.  In general the affine connection $\Gamma$ on the spacetime manifold $M$ has non-vanishing torsion. It might be the case that torsion plays an important role in cosmology, but in this work we concentrate on the case where torsion is not present; i.e., we impose $T^a_{\mu\nu}=0$.  It was demonstrated in \cite{Gegenberg:2015gma} that the torsion-free condition is preserved under the gauge transformations (\ref{eq:gauge 2}).  Another assumption concerns the dependence of the matter action on the gauge potential.  Specifically, we assume that the matter action is a functional of the metric $g_{\alpha\beta}$, the field $l^{a}_{\alpha}$ gauging special conformal transformations, and matter fields (generically denoted by $\psi$) only:
\begin{equation}
	S_\text{m} = S_\text{m}[g_{\alpha\beta},l^{a}_{\alpha},\psi].
\end{equation}
More general types of matter-gauge potential coupling are discussed in \cite{Gegenberg:2015gma}.  Finally, in the full theory derived from (\ref{eq:action}) there is an antisymmetric tensor,
\begin{equation}
	 \f_{\alpha\beta} = \tfrac{1}{2} \eta_{ab} e^{a}_{[\alpha} l_{\beta]}^{b},
\end{equation}
appearing in the field equations that satisfies Maxwell-like equations for the electromagnetic field strength.  Since our primary interest is cosmology below, we expect such a tensor would be ruled out by isotropy and homogeneity and hence we set $\f_{\alpha\beta} = 0$.  We note that it is easily confirmed that the vanishing of $\f_{\alpha\beta}$ is a gauge invariant condition:  Under the transformations (\ref{eq:gauge 2}), we have:
\begin{equation}
	\delta \f_{\alpha\beta} = -2 \nabla_{[\alpha} \nabla_{\beta]} \Omega = 0.
\end{equation}

Variation of the action (\ref{eq:action}) with respect to the gauge potential under these assumptions yields the equations of motion:
\begin{subequations}\label{eq:EOMs}
\begin{align}\label{eq:Bach EOM}
	0 & = B^{\alpha\nu} +\tfrac{1}{16} g_\text{YM}^{2}  \mathscr{T}^{\alpha\nu} - \nabla_{\mu}\nabla^{[\nu}\bar{a}^{\mu]\alpha} - \mathscr{Q}^{\alpha\nu},\\
	0 & = \nabla^{\alpha} a_{\alpha\beta}\label{eq:a conserve}, \\
	0 & = \nabla_{\beta}a.\label{eq:a trace}
\end{align}
\end{subequations}
Here and below, the vanishing of the torsion implies that $\nabla_{\alpha}$ is the ordinary covariant derivative operator as defined from the Levi-Civita connection.  Also, $a_{\alpha\beta}$ describes matter coupling to $l^{a}_{\alpha}$ while $\mathscr{T}_{\alpha\beta}$ is the ordinary stress-energy tensor:
\begin{gather}
	a^{\mu\nu} =  \frac{g_\text{YM}^{2}}{4\sqrt{-g}} \frac{\delta(\sqrt{-g} \mathcal{L}_\text{m})}{\delta l^{b}_{\nu}} e^{b\mu}, \nonumber \\ \mathscr{T}_{\mu\nu} = -\frac{2}{\sqrt{-g}} \frac{\delta(\sqrt{-g} \mathcal{L}_\text{m})}{\delta g^{\mu\nu}}. \label{eq:a and b def}
\end{gather}
The other quantities appearing in (\ref{eq:EOMs}) are given by:
\begin{subequations}
\begin{align}
\mathscr{Q}^{\alpha\nu} & = \tfrac{1}{2} a_{\lambda\mu} C^{\alpha\lambda\mu\nu} - \tfrac{1}{8} \tau^{\alpha\nu}-\tfrac{1}{2}(2S_{\lambda\mu} - \bar{a}_{\lambda\mu}) \nonumber \\ & \times (g^{\lambda[\mu} \bar{a}^{\nu]\alpha} - g^{\alpha[\mu} \bar{a}^{\nu]\lambda}), \\
B_{\mu\nu} & = -\nabla^{\alpha} \nabla_{\alpha} S_{\mu\nu} + \nabla^{\alpha} \nabla_{\mu} S_{\alpha\nu} + C_{\mu\alpha\nu\beta} S^{\alpha\beta}, \\
S_{\alpha\beta} & = \tfrac{1}{2}(R_{\alpha\beta} - \tfrac{1}{6} R g_{\alpha\beta}), \\
\tau^{\rho\sigma} & =  -4 C^{\alpha\beta\gamma\rho}C_{\alpha\beta\gamma}{}^\sigma + g^{\rho\sigma} C^{\alpha\beta\gamma\delta} C_{\alpha\beta\gamma\delta} , \\
\bar{a}_{\alpha\beta} & = a_{\alpha\beta} - \tfrac{1}{6} g_{\alpha\beta} a, \\
a & =a^{\alpha}{}_{\alpha}.
\end{align}
\end{subequations}
Here, $C_{\alpha\beta\gamma\delta}$ is the Weyl tensor, while $S_{\alpha\beta}$ and $B_{\mu\nu}$ are the Schouten and Bach tensors.  Finally, by taking the divergence of (\ref{eq:Bach EOM}), we find that the stress-energy tensor is conserved as usual: $\nabla^{\alpha} \mathscr{T}_{\alpha\beta}=0$.

\section{Friedmann-Robertson-Walker cosmology}\label{sec:frw}

\subsection{The Friedmann equation}\label{sec:friedmann}

The goal of this section is to study the evolution of a spatially homogenous and isotropic spacetime in our model.  We therefore assume the Friedmann-Robertson-Walker (FRW) line element:
\be
ds^{2}=-dt^{2}+A^{2}\left( \frac{dr^{2}}{1-kr_{0}^{2}/r^{2}}+r^{2}d\theta^{2}+r^{2}\sin^{2} \theta \, d\phi^{2}\right),
\ee
where $A = A(t)$ is the scale factor, $r_{0}$ is a constant with the dimension of length, and $k=0,+1,-1$ for flat, 3-sphere and 3-hyperboloid spatial geometries, respectively. 

Due to the isotropy and homogeneity of the spacetime, the symmetric matter source $a_{\mu\nu}$ must take the form:
\be
\label{eq: aperfect}
a^{\mu\nu} =\left(\xi_1+\xi_2\right)u^{\mu}u^{\nu}+\xi_2g^{\mu\nu},
\ee
where $u^{\alpha} \di_{\alpha} = \di_{t}$.  This is algebraically identical to the stress-energy tensor of a perfect fluid, but we caution that $a_{\mu\nu}$ should not be interpreted in this way:  As seen in (\ref{eq:a and b def}), this tensor arises from the variation of the matter action with respect to $l^{a}_{\alpha}$, not from the variation with respect the metric.  Also, the parameters $\xi_1$ and $\xi_2$ appearing in (\ref{eq: aperfect}) do not have the dimensions of density and pressure; rather, they have dimensions of $(\text{mass})^{2}$.  Substituting the trace of (\ref{eq: aperfect}) into (\ref{eq:a conserve}) and (\ref{eq:a trace}) and solving yields
 \be
 \label{eq:a soln}
\xi_1= -\frac{\Pi}{A^{4}} - \Lambda, \quad \xi_2=-\frac{\Pi}{3A^{4}}+ \Lambda,
 \ee
where $\Lambda$ and $\Pi$ are constants of integration. 

For any homogeneous and isotropic spacetime the Bach and Weyl tensors vanish identically, and therefore $(\ref{eq:Bach EOM})$ reduces to
\begin{multline}
\label{eq: beq}
\tfrac{1}{16} g_\text{YM}^{2} \mathscr{T}^{\alpha\nu} = \nabla_{\mu}\nabla^{[\nu}\bar{a}^{\mu]\alpha} -\tfrac{1}{2}(2S_{\lambda\mu} - \bar{a}_{\lambda\mu}) \\ \times (g^{\lambda[\mu} \bar{a}^{\nu]\alpha} - g^{\alpha[\mu} \bar{a}^{\nu]\lambda} ) .
\end{multline}
We fix the ``ordinary'' matter content of the universe to be non-interacting pressureless dust and radiation as in $\Lambda$CDM:
\be
\mathscr{T}^{\mu\nu} = \mathscr{T}_{\mu\nu}^{(\text{m})} + \mathscr{T}_{\mu\nu}^{(\text{r})}, 
\ee
with
\begin{gather}
\mathscr{T}_{\mu\nu}^{(\text{m})} =  \rho_\text{m} u_{\mu} u_{\nu}, \quad \mathscr{T}_{\mu\nu}^{(\text{r})} =  (\rho_\text{r}+p_\text{r}) u_{\mu} u_{\nu} + p_\text{r} g_{\mu\nu}, \nonumber \\  p_\text{r} = \rho_\text{r}/3.
\end{gather}
Demanding that each matter source is separately conserved ($\nabla^{\mu} \mathscr{T}^{(\text{m})}_{\mu\nu}=\nabla^{\mu} \mathscr{T}^{(\text{r})}_{\mu\nu} =0$) yields:
\begin{subequations}
\begin{align}
\dot{\rho}_\text{m}+3H\rho_{\text{m}} & =0 & \Rightarrow & & \rho_\text{m}& ={\rho_{\text{m},0}}{A^{-3}}, \\
\dot{\rho}_\text{r}+4H \rho_{\text{r}} &  =0  & \Rightarrow & & \rho_\text{r} & = {\rho_{\text{r},0}}{A^{-4}}.
\end{align}
\end{subequations}
where $\rho_{\text{m},0}$ and $\rho_{\text{r},0}$ are constants, we use an overdot to denote $d/dt$, and we have defined the Hubble parameter $H = \dot{A}/A$. The $(00)$ component of (\ref{eq: beq}) then yields the Friedmann equation
\be
\label{eq:= friedmann1}
H^{2}=\frac{g_{\textrm{YM}}^{2}}{8} \left( \frac{\rho_{\text{m}}+\rho_{\text{r}}}{\Lambda+\frac{\Pi}{A^{4}}} \right) -\frac{k}{r_{0}^{2}A^{2}}+\frac{\Lambda}{3}-\frac{\Pi}{3A^{4}},
\ee
where we have used (\ref{eq: aperfect}) and (\ref{eq:a soln}).\footnote{The spatial components of (\ref{eq: beq}) yield an equation for $\dot{H}$ that can be derived from the formula already presented.}  In the late-time limit $A^{4} \gg |\Pi/\Lambda|$, we obtain
\be
\label{eq: friedmancosmo}
H^{2}\approx\frac{\rho_{\text{m}}+\rho_{\text{r}} }{3M_{\text{Pl}}^{2}}- \frac{k}{r_{0}^{2}A^{2}} +\frac{\Lambda}{3},
\ee
where we have identified the Planck mass as:
\be
\label{eq: gym}
M_{\text{Pl}}^{2}=\frac{8}{3}\frac{\Lambda}{g_{\textrm{YM}}^{2}}.
\ee
Equation $(\ref{eq: friedmancosmo})$ is the same as the Friedmann equation in $\Lambda$CDM provided that we interpret the constant of integration $\Lambda$ as the cosmological constant.  Probes of the expansion history of the late time universe gives us the order of magnitude of $\Lambda$:
\be
\Lambda \sim \frac{(10^{-3} \, \text{eV})^{4}}{M_\text{Pl}^{2}} \sim (10^{-33} \, \text{eV})^{2}.
\ee
This in turn fixes the size of Yang-Mills coupling constant to be $g^{2}_{\textrm{YM}} \sim 10^{-120}$.

Before moving on, we make a few remarks about the interpretation of the Friedmann equation (\ref{eq:= friedmann1}).  This equation can be rewritten in a form more familiar from general relativity if one introduces a time-varying Newton's constant and a ``dark radiation'' field with density $\propto -\Pi$:
\be
\label{eq:= friedmann3}
H^{2}=\frac{8\pi G_\text{eff}(A)}{3} \left( \rho_{\text{m}}+\rho_{\text{r}}  \right) -\frac{k}{r_{0}^{2}A^{2}}+\frac{\Lambda}{3}-\frac{\Pi}{3A^{4}},
\ee
where
\begin{equation}
G_\text{eff}(A) = \frac{3g_{\textrm{YM}}^{2}}{64 \pi} \left( \frac{A^{4}}{\Lambda A^{4} + \Pi} \right)  = \frac{1}{8\pi M_\text{Pl}^{2}} \left( \frac{A^{4}}{A^{4} + \Pi/\Lambda } \right).	
\end{equation}
We see that the effective Newton constant decreases with decreasing $A$; i.e., the force of gravity is weaker in the past.  As seen in section \ref{sec:dynamics}, this screening of the gravitational field sourced by ordinary matter will have important consequences for early universe dynamics.

We also note that if $\Pi = 0$, we recover the Friedmann equation of general relativity exactly.  Indeed, if $\Pi = 0$ we have
\begin{equation}
	a_{\alpha\beta} = \Lambda g_{\alpha\beta},
\end{equation}
which when substituted into equation (\ref{eq: beq}) yields:
\begin{equation}
	G_{\alpha\beta} + \Lambda g_{\alpha\beta} = \frac{3g_\text{YM}^{2}}{8\Lambda} \mathscr{T}_{\alpha\beta}.
\end{equation}
Here, $G_{\alpha\beta}$ is the Einstein tensor, so this is equivalent to the field equations of general relativity with a cosmological constant provided we identify the Planck mass as in (\ref{eq: gym}).

\subsection{Cosmological dynamics}\label{sec:dynamics}

It is useful to write the Friedmann equation in terms of the same density parameters used to describe the $\Lambda$CDM model:
\begin{gather}
\Omega_{\text{m}}=\frac{\rho_{\text{m},0}}{3M_{\text{Pl}}^{2}H_{0}^{2}}, \quad \Omega_{\text{r}}= \frac{\rho_{\text{r},0}}{3M_{\text{Pl}}^{2}H_{0}^{2}}, \nonumber \\ \Omega_{\Lambda}=\frac{\Lambda}{3H_{0}^{2}},\quad \Omega_{\text{k}}=-\frac{k}{r_{0}^{2}H_{0}^{2}},
\end{gather}
where we have assumed that $A=1$ and $H=H_{0}$ at the present epoch. We also define a dimensionless density parameter for the ``dark radiation'' (i.e.\ the integration constant $\Pi$):
\be
\label{eq: omegat}
\quad \Omega_{\Pi}=\frac{\Pi}{3H_{0}^{2}}.\ee
Note that since observations imply that $\Lambda \sim H_{0}^{2}$, we have that $\Omega_{\Pi} \sim \Pi/\Lambda$; i.e., it is roughly the ratio of the two constants appearing in our solution for $a_{\alpha\beta}$.  In terms of these, the Friedmann equation (\ref{eq:= friedmann1}) becomes
\be
\label{eq: friedm}
\frac{H^{2}}{H_{0}^{2}}=\frac{\Omega_{\Lambda}\Omega_{\text{m}}}{A^{3}(\Omega_{\Lambda}+\frac{\Omega_{\Pi}}{A^{4}})}+\frac{\Omega_{\Lambda}\Omega_{\text{r}}}{A^{4}(\Omega_{\Lambda}+\frac{\Omega_{\Pi}}{A^{4}})}+\frac{\Omega_{\text{k}}}{A^{2}}+\Omega_{\Lambda}-\frac{\Omega_{\Pi}}{A^{4}},
\ee
where we have made use of (\ref{eq: gym}).  Evaluating this at the present epoch (when $A=1$ and $H = H_{0}$) yields a constraint amongst the density parameters:
\be
\label{eq: omega0}
1=\frac{\Omega_{\Lambda}(\Omega_{\text{m}}+\Omega_\text{r}) }{\Omega_{\Lambda}+\Omega_{\Pi}}+\Omega_{\text{k}}+\Omega_{\Lambda}-\Omega_{\Pi}.
\ee
Note that if $|\Omega_{\Pi}|\ll1$ we recover the standard $\Lambda$CDM relation
\be
1=\Omega_{\text{m}}+\Omega_{\text{r}}+\Omega_{\text{k}}+\Omega_{\Lambda}.
\ee

In order to qualitatively analyze the cosmological dynamics, it is useful to rewrite the Friedmann equation as the equation of motion of a zero-energy particle moving in a one-dimensional effective potential:
\begin{equation}\label{eq:EOM}
	\frac{1}{2} \left( \frac{dA}{d\tau} \right)^{2} + V_\text{eff}(A) = 0,
\end{equation}
where we have defined $\tau = H_{0} t$ and
\begin{multline}
\label{eq: vomega}
V_{\text{eff}}(A)=-\frac{\Omega_{\Lambda}\Omega_{\text{m}}}{2A(\Omega_{\Lambda}+\frac{\Omega_{\Pi}}{A^{4}})}-\frac{\Omega_{\Lambda}\Omega_{\text{r}}}{2A^{2}(\Omega_{\Lambda}+\frac{\Omega_{\Pi}}{A^{4}})} \\ -\frac{\Omega_{\text{k}}}{2}-\frac{\Omega_{\Lambda}A^{2}}{2}+\frac{\Omega_{\Pi}}{2A^{2}}.
\end{multline}
We note that (\ref{eq: omega0}) can be used to eliminate $\Omega_{\text{k}}$ in either (\ref{eq: friedm}) or (\ref{eq: vomega}).  The utility of the Friedmann equation written as (\ref{eq:EOM}) is that we can immediately conclude that all values of the scale factor with $V_\text{eff}(A) > 0$ are classically forbidden, and we can obtain the acceleration of the universe in a given epoch from $\ddot{A} = - V_\text{eff}'(A)$.  It is also of interest to define the ``slow-roll'' parameter
\be
\label{eq: epsilon}
\epsilon_\text{H}=-\frac{\dot{H}}{H^{2}} = 1 - \frac{\ddot{A}}{H^{2}A} =1-\frac{A}{2}\frac{V'_{\text{eff}}(A)}{V_\text{eff}(A)}.
\ee
This is a direct measure of the rate of change of the Hubble parameter.  Using these quantities, we obtain three equivalent conditions for the Universe to be accelerating:
\begin{equation}
	\ddot{A} > 0 \quad \Leftrightarrow \quad \epsilon_\text{H} < 1 \quad \Leftrightarrow \quad V_\text{eff}'(A)<0.
\end{equation}

\begin{figure*}
\begin{center}
\includegraphics[width=\textwidth]{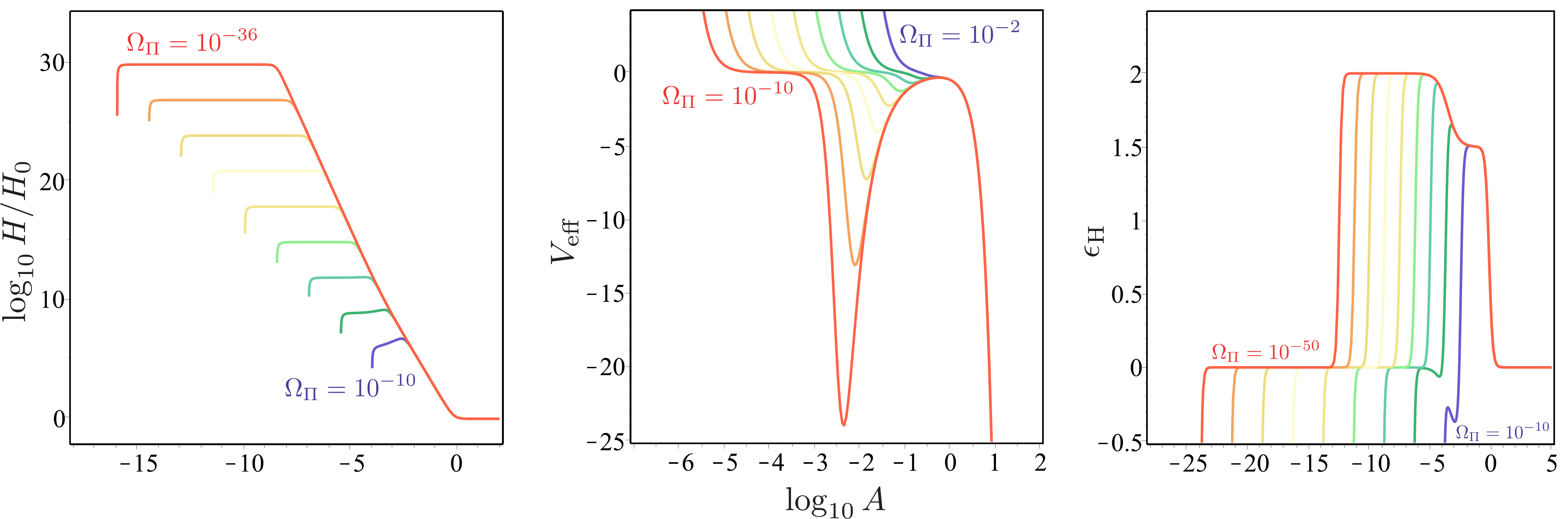}
\end{center}
\caption{Hubble parameter $H$ (left), effective potential $V_\text{eff}$ (centre), and slow-roll parameter $\epsilon_\text{H}$ (right) as functions of scale factor.  Here, we have taken $(\Omega_\text{m},\Omega_\text{r},\Omega_{\Lambda})=(0.27,8.24\times10^{-5},0.73)$.  There is a cosmological bounce at early times when $H=0$, $V_\text{eff}=0$, and $\epsilon_\text{H}\rightarrow -\infty$.   This bounce is followed by a period of quasi de-Sitter acceleration when $H \approx$ constant, $V \propto -A^{2}$, and $\epsilon_\text{H} \approx 0$.  The early time acceleration ends when $V_\text{eff}$ switches from decreasing to increasing and is followed by epochs of radiation, matter, and late time acceleration similar to $\Lambda$CDM.}\label{fig:potential}
\end{figure*}
Now, in order to be consistent with late times probes of cosmological expansion (such as Supernovae of Type IA), we demand that the Friedman equation (\ref{eq: friedm}) reduce down to the $\Lambda$CDM form when $A\gtrsim 1$.  This implies that $|\Omega_{\Pi}| \ll 1$.  Furthermore, to avoid a singularity in the the Friedmann equation for finite $A>0$, we will assume that $\Omega_{\Pi} > 0$.\footnote{This is not a necessary assumption, and it would be interesting to consider the $\Omega_\Pi < 0$ case in future work.}  Given that we recover $\Lambda$CDM for $A \gtrsim 1$, we expect that the other density parameters will take on their concordance values \cite{carroll2007introduction}:
\be\label{eq:observational values}
\Omega_{\text{m}} = 0.27\pm0.04,\,\,\, \Omega_{\Lambda} = 0.73\pm0.04, \,\,\, \Omega_{\text{r}}\simeq8.24\times10^{-5}.
\ee

In Figure \ref{fig:potential}, we plot the Hubble parameter, effective potential and slow-roll parameter as functions of the scale factor assuming the central values of cosmological parameters in (\ref{eq:observational values}).  Since values of the scale factor for which $V_\text{eff}(A)>0$ are classically forbidden, there will be an early Universe ``big bounce'' that occurs when $V_{\text{eff}}(A)=0$ and there is no big bang singularity in our model.  The replacement of the big bang with a big bounce in this model is a direct consequence of the weakening of the gravitational field sourced by ordinary matter in the early universe (c.f.\ equation \ref{eq:= friedmann1}).  Essentially, strong gravitational forces implied by high densities are mitigated by the reduction of $G_\text{eff}$ in the distant past, which allows the universe to escape an initial singularity.   

\begin{figure}
\begin{center}
\includegraphics[width=\columnwidth]{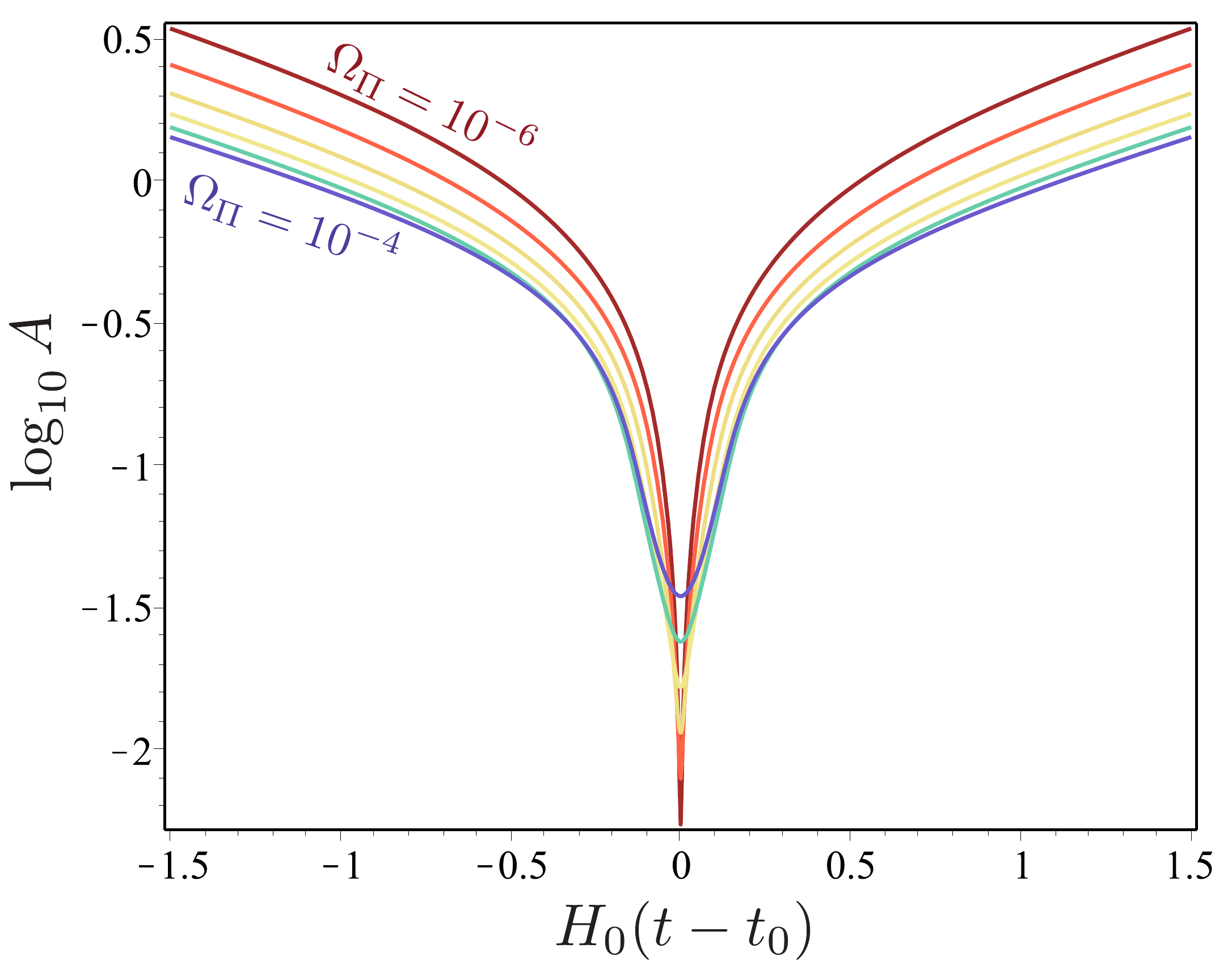}
\end{center}
\caption{Numeric solutions for the scale factor $A$ assuming $(\Omega_\text{m},\Omega_\text{r},\Omega_{\Lambda})=(0.27,8.24\times10^{-5},0.73)$.  All simulations show a bounce at time $t=t_{0}$.  The scale factor at the bounce increases with increasing $\Omega_{\Pi}$.}\label{fig:scale factor}
\end{figure}

Immediately after this cosmological bounce there is a phase of nearly dS early-time acceleration.  The universe undergoes two further transitions where $V_\text{eff}'(A) = 0$:  The first transition marks when the acceleration in the early Universe ends and the radiation dominated epoch starts, and the second transition occurs in the late Universe when matter domination ends and the second acceleration epoch starts. The latter is consistent with the observed late-time acceleration of the Universe.  We give an example of numerical solutions of (\ref{eq:EOM}) for the scale factor in Figure \ref{fig:scale factor}, which clearly demonstrates the existence of a bounce in the early Universe.

As mentioned above, in order to recover an acceptable late-time cosmology, we must have that $|\Omega_{\Pi}| \ll 1$.  Let us assume that $0<\Omega_{\Pi} \ll \Omega_{\Lambda}$, and hence obtain the following approximate form of the potential:
\begin{multline}
	V_\text{eff}(A) \approx - \frac{1}{2} \left( \frac{\Omega_\text{m}}{A} + \frac{\Omega_\text{r}}{ A^{2}} \right) \left( 1 + \frac{\Omega_{\Pi}}{\Omega_{\Lambda} A^{4}} \right)^{-1} \\ + \frac{\Omega-1}{2} + \frac{\Omega_{\Pi}}{2A^{2}} - \frac{\Omega_{\Lambda} A^{2} }{2},
\end{multline}
where we have defined
\begin{equation}
	\Omega = \Omega_\text{m} + \Omega_\text{r} + \Omega_{\Lambda},
\end{equation}
as in standard $\Lambda$CDM cosmology.  By making further assumptions on the size of $A$ and preforming some straightforward analysis, we can write $V_\text{eff}$ in various epochs:
\begin{equation}\label{eq:approx V}
	V_\text{eff}(A) \approx \frac{\Omega-1}{2} -  \frac{1}{2} \begin{cases}
	 \Omega_\text{r}\Omega_{\Lambda}  \Omega^{-1}_{\Pi} A^{2} & A_{1} \ll A \ll A_{2}, \\
	 \Omega_\text{r} A^{-2}, & A_{2} \ll A \ll A_{3}, \\
	 \Omega_\text{m} A^{-1}, & A_{3} \ll A \ll A_{4}, \\
	 \Omega_{\Lambda} A^{2}, & A_{4} \ll A,
	\end{cases}
\end{equation}
where we have defined
\begin{align}\nonumber
	A_{1} & =  \left[ \frac{\Omega_{\Pi}}{\Omega_{\Lambda}\Omega_\text{r}} \left( \frac{\Omega-1}{2} + \sqrt{ \frac{(\Omega-1)^{2}}{4} +\Omega_{\Lambda} \Omega_\text{r} } \,   \right) \right]^{1/2}, \\ A_{2} & = \left( \frac{\Omega_{\Pi}}{\Omega_{\Lambda}} \right)^{1/4},  \quad A_{3}  = \frac{\Omega_\text{r}}{\Omega_\text{m}},  \quad A_{4}  = \left( \frac{\Omega_\text{m}}{\Omega_{\Lambda}} \right)^{1/3}. \label{eq:epochs}
\end{align}
To obtain (\ref{eq:approx V}), we have further assumed that
\begin{gather}\nonumber
	\Omega = \mathcal{O}(1), \quad \Omega_{\Lambda}= \mathcal{O}(1), \quad \Omega_\text{m} = \mathcal{O}(1), \\ \Omega_\text{r} \ll \Omega_\text{m}, \quad \Omega_{\Pi} \ll \Omega_\text{r}^{2} \ll \Omega_\text{r},
\end{gather}
in order to ensure the hierarchy $A_{1} \ll A_{2} \ll A_{3} \ll A_{4}$.  (These assumptions are all consistent with the observational values quoted above.)  We can name the various epochs (\ref{eq:approx V}) in analogy to the behaviour of $V_\text{eff}$ in standard cosmology:
\begin{center}
\begin{tabular}{rl}
early quasi-dS acceleration: & $A_{1} \ll A \ll A_{2}$  \\
radiation domination: & $A_{2} \ll A \ll A_{3}$ \\
matter domination: & $A_{3} \ll A \ll A_{4}$ \\
late quasi-dS acceleration: & $A_{4} \ll A$.
\end{tabular}  
\end{center}
In particular, if we assume $\Omega \approx 1$ then we find that in the ``early quasi-dS acceleration'' phase
\begin{equation}
	A \approx \exp [\Omega_{r}^{1/2}\Omega_{\Lambda}^{1/2} \Omega_{\Pi}^{-1/2} H_{0}(t-t_{0}) ], \quad  A_{1} \ll A \ll A_{2};
\end{equation}
i.e., we have exponential expansion ($t_{0}$ is an integration constant).

\begin{figure}
\begin{center}
\includegraphics[width=0.4\textwidth]{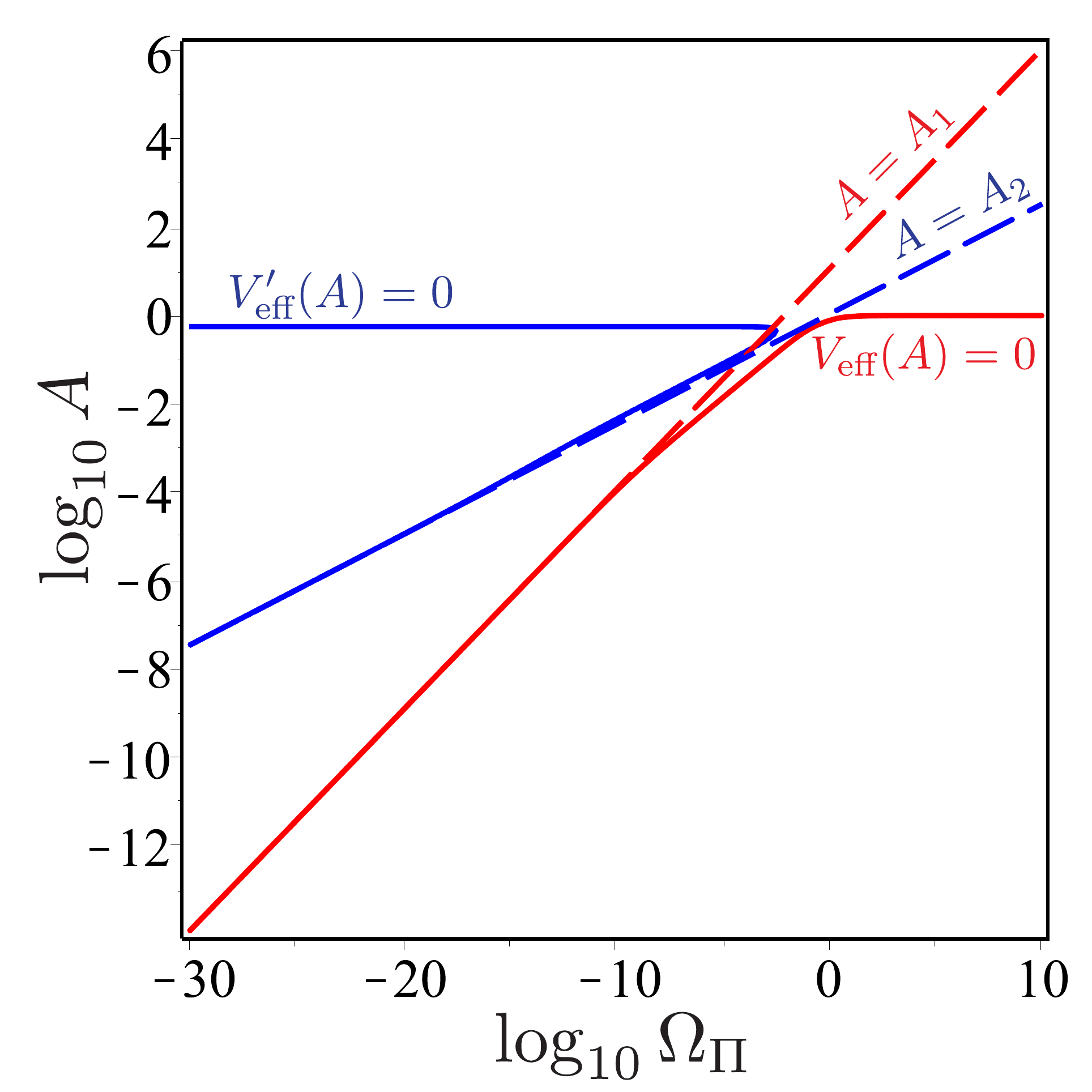}
\end{center}
\caption{Curves $V_\text{eff}(A) = 0$, $V'_\text{eff}(A) = 0$, $A=A_{1}$ and $A=A_{2}$ in the $(\Omega_{\Pi},A)$ plane with $(\Omega_\text{m}, \Omega_\text{r}, \Omega_{\Lambda})=(0.27,8.24\times10^{-5},0.73)$.  We see that for $0 < \Omega_{\Pi} \ll 1$ the $V_\text{eff}(A) =0$ and $A=A_{1}$ curves coincide, while the $V'_\text{eff}(A) =0$ and $A=A_{2}$ curves coincide.  This is an explicit confirmation that the bounce occurs at $A \approx A_{1}$ and the quasi-dS inflation ends when $A \approx A_{2}$ for these parameters and $0<\Omega_{\Pi} \ll 1$.}\label{fig:curves}
\end{figure}

From figure \ref{fig:potential}, we expect the early acceleration phase to be preceeded by a cosmological bounce that occurs when $V_\text{eff}(A) = 0$.  By performing a 3-term Taylor series expansion of (\ref{eq: vomega}) about $A=0$ and working to leading order in $\Omega_{\Pi}$, we find that
\begin{equation}
	V_\text{eff}(A_{1}) \approx 0, \quad 0 < \Omega_{\Pi} \ll 1,
\end{equation}
where $A_{1}$ is given by (\ref{eq:epochs}).  That is, the bounce occurs at $A \approx A_{1}$ when $\Omega_{\Pi}$ is small and positive.  On the other hand, the transition from early time acceleration to radiation domination at $A \approx A_{2}$ occurs when the potential switches from a decreasing to increasing function of $A$.  Therefore, we also expect
\begin{equation}
	V'_\text{eff}(A_{2}) \approx 0, \quad 0 < \Omega_{\Pi} \ll 1,
\end{equation}
where $A_{2}$ is given by  (\ref{eq:epochs}).  To test these approximations, we can plot the curves $V_\text{eff}(A) = 0$, $V'_\text{eff}(A) = 0$, $A=A_{1}$ and $A=A_{2}$ in the $(\Omega_{\Pi},a)$ plane with $(\Omega_\text{m}, \Omega_\text{r}, \Omega_{\Lambda})$ held constant.  An example of such a plot is given in Figure \ref{fig:curves}.

Given formulae for $A_{1}$ and $A_{2}$, we can estimate how many e-folds $N$ of exponential expansion occur after the bounce:
\begin{multline}
	N = \ln \frac{A_{2}}{A_{1}} = - \frac{1}{4} \ln \Omega_{\Pi} + \frac{1}{4} \ln \Omega_{\Lambda} + \frac{1}{2} \ln \Omega_\text{r} \\ - \frac{1}{2} \ln
	 \left( \frac{\Omega-1}{2} + \sqrt{ \frac{(\Omega-1)^{2}}{4} +\Omega_{\Lambda} \Omega_\text{r} } \,   \right).
\end{multline}
We see from this that we can make $N$ arbitrarily large by selecting $\Omega_{\Pi}$ to be very small.  If we take cosmological parameters as their central values in (\ref{eq:observational values}), then we have
\begin{equation}
	N \sim 60 - \frac{1}{4} \ln \frac{\Omega_{\Pi}}{10^{-109}} \sim 66 - \frac{1}{4} \ln \frac{\Omega_{\Pi}}{g_\text{YM}^{2}}.
\end{equation}
We also note that the Hubble scale during this early ``inflationary'' period is also fixed by $\Omega_{\Pi}$:
\begin{equation}
	H_{\inf} \approx \Omega_{r}^{1/2}\Omega_{\Lambda}^{1/2} \Omega_{\Pi}^{-1/2}  H_{0}.
\end{equation}
This is commonly characterized by the energy scale during inflation:
\begin{equation}
	E_\text{inf} = (3M_\text{Pl}^{2} H_\text{inf}^{2})^{1/4} \approx 3^{1/4} \Omega_\text{r}^{1/4} \Omega_{\Lambda}^{1/4} \Omega_{\Pi}^{-1/4} M_\text{Pl}^{1/2} H_{0}^{1/2}.
\end{equation}
Again taking central values for the usual density parameters and $H_{0} \sim 10^{-33} \, \text{eV}$, we find
\begin{equation}
	E_\text{inf} \sim 10^{15} \, \text{GeV} \left( \frac{\Omega_{\Pi}}{ 10^{-109} } \right)^{-1/4} \sim 5 \times 10^{17} \, \text{GeV} \left( \frac{\Omega_{\Pi}}{ g_\text{YM}^{2} } \right)^{-1/4}.
\end{equation}
We note the relationship between $N$ and $E_\text{inf}$ in this model
\begin{equation}
	N \sim 60 + \ln \left( \frac{ E_\text{inf}} {10^{15} \, \text{GeV}} \right),
\end{equation}
again assuming central values for $(\Omega_\text{m},\Omega_\text{r},\Omega_\Lambda)$.

Finally, for this model to accurately reproduce observed light element abundances, we require that the cosmological expansion history from big bang nucleosynthesis onwards be not significantly different than that of standard $\Lambda$CDM.  This can be guaranteed if we have $A_{2} \ll A_\text{BBN}$, where $A_\text{BBN}$ is the scale factor at big bang nucleosynthesis.  Using standard formulae, this condition can be re-written as
\begin{equation}
	\Omega_{\Pi} \ll 2 \times 10^{-35} \left( \frac{T_\text{BBN}}{100\,\text{keV}} \right)^{-4},
\end{equation}
where $T_\text{BBN}$ is the radiation temperature at big bang nucleosynthesis.

\section{Discussion}\label{sec:discussion}

In this paper we have considered cosmological solutions of a gauge theory of gravity.  The action of our model resembles that of a Yang-Mills theory with gauge group SO(4,2); i.e.\ the conformal group of Minkowski space.  The metric and connection of the spacetime manifold are identified from various components of the gauge potential.  The ensuing gravitational theory is in general rather rich and complex, but we made a number of simplifying assumptions to aid our analysis.  For example, the full theory admits manifolds with torsion and matter which couples to the gauge potential in exotic ways.  In this paper, we considered torsion-free solutions and retained only matter directly coupled to the metric (as in general relativity) and matter coupled to fields $l^{a}_{\alpha}$ gauging special conformal transformations.  Our model is invariant under local conformal (Weyl) transformations.

When we specialized to isotropic and homogeneous spacetimes, we found that the contribution to the field equations of matter coupled to $l^{a}_{\alpha}$ is highly constrained.  Indeed, the tensor $a_{\alpha\beta}$ encoding this contribution is completely determined by two integration constants: $\Lambda$ and $\Pi$.  We derived the Friedman equation governing the dynamics and deduced that at late times the model reduces down to the standard $\Lambda$CDM cosmology with $\Lambda$ playing the role of the cosmological constant.  It is worth noting that the cosmological constant in our model was not put into the action by hand, as in $\Lambda$CDM, rather it is generated dynamically from the matter coupled to $l^{a}_{\alpha}$.

If the other constant $\Pi$ in the solution for $a_{\alpha\beta}$ is set to zero, we recover $\Lambda$CDM exactly for all times.  However, if it is not zero there are fascinating repercussions in the early universe.  If $\Pi > 0$, the big bang of general relativity is replaced by a cosmological bounce.  Furthermore if $0 < \Pi \ll H_{0}^{2}$, then the bounce is followed a period of quasi-dS acceleration.  That is, there exists a period of slow-roll inflation in the early universe.  This inflationary period can be made arbitrary long by selecting $\Pi$ to be arbitrarily small.  The physical reason for these effects is that the effective Newton constant mediating the gravitational force exerted by ordinary matter (i.e.\ dust and radiation) becomes small in the past, allowing for a bounce.

To summarize, we have presented a theory of gravity whose cosmological solutions are free of singularities, and which incorporate quasi-dS epochs of acceleration in the early and late universe.  One may be concerned about the naturalness of such a theory.  \emph{A priori}, our cosmological solutions involve one dimensionless constant appearing in the action $g^{2}_\text{YM}$, and two dimensionful constants of integration $\Lambda$ and $\Pi$.  We fixed $\Lambda$ by comparing to observations of late time acceleration.  The Yang-Mills coupling was then fixed by requiring the late time Friedmann equation have the correct dependence on the Planck mass.  Since there is a large hierarchy between the Planck and dark energy scales, this yielded a small Yang-Mills coupling $g^{2}_\text{YM} \sim 10^{-120}$.  In order to recover acceptable late time cosmology, we required $\Pi \ll \Lambda$, which implies the most ``natural'' nonzero value for $\Pi$ is
\begin{equation}
	\Pi \sim g_\text{YM}^{2} \Lambda \quad \Rightarrow \quad \Omega_{\Pi} \sim g_\text{YM}^{2}.
\end{equation}
With this choice, the early time dS-phase involves $\sim 66$ e-folds of exponential expansion (which is sufficient to explain the homogeneity of the cosmic microwave background) at an energy scale of $\sim 5\times 10^{17} \, \text{GeV}$ (which implies high temperature inflation).  Furthermore, this value of $\Pi$ will yield an expansion history consistent with big bang nucleosynthesis.  Therefore, just as in $\Lambda$CDM, our model does involve one unnaturally small number forced upon us by the observed hierarchy between the Planck mass and cosmological constant; the other constant $\Pi$ can take on a natural value and still generate an acceptable cosmological model. 

In the future, this model needs to be rigorously compared with observations.  By comparing the predictions of the modified Friedmann equation (\ref{eq: friedm}) with probes of the expansion history (such as type IA supernove), we can obtain direct bounds on $\Omega_{\Pi} \sim \Pi/\Lambda$.  Perhaps more importantly, as shown in \cite{Gegenberg:2015gma}, matter perturbations in this model can source long-range gravitational forces.  This means that the dynamics of cosmological perturbations may be significantly different from general relativity, which could lead to definitive observational tests of the model; both in the late universe via observations of large scale structure and in the early universe via the quantum generation of fluctuations during inflation.   (Primordial perturbations in models with a cosmological bounce have been considered in \cite{Cai:2008ed,Cai:2008qb,Cai:2012va,Xia:2014tda,Cai:2014bea}.)   Finally, the role of torsion in this model is interesting at both the background and perturbative level, and needs to be explored in greater detail.

\begin{acknowledgments}

SR would like to thank Jonathan Ziprick for useful discussions. We are supported by National Sciences and Engineering Research Council of Canada (NSERC).

\end{acknowledgments}

\bibliographystyle{apsrev4-1}
\bibliography{gravity_yang_mills}

\end{document}